\documentstyle[12pt,fleqn]{article}
\begin{document}
\pagestyle{empty}

\title{ Relativistic Quantum Transport Theory \\
        for Electrodynamics\thanks{Supported
        in part by DFG, GSI and BMFT.}}
\author{P. Zhuang\thanks{On leave
        from Hua-Zhong Normal University, Wuhan, China }\ \
        and \ U. Heinz\\ Institut f\"ur Theoretische Physik,
        Universit\"at Regensburg,\\ D-93040 Regensburg, Germany }

\maketitle

\begin{abstract}

\baselineskip 0.6cm

We investigate the relationship between the covariant and the
three-dimensional (equal-time) formulations of quantum kinetic theory.
We show that the three-dimensional approach can be obtained as the
energy average of the covariant formulation. We illustrate this
statement in scalar and spinor QED. For scalar QED we derive Lorentz
covariant transport and constraint equations directly from the
Klein-Gordon equation rather than through the previously used
Feshbach-Villars representation. We then discuss the relation to their
equal-time version. We make a semiclassical expansion in $\hbar$ and
obtain Vlasov-type transport equations for the zeroth and first order
of the scalar Wigner operator. We then consider pair production in a
spatially homogeneous but time-dependent electric field and show that
the pair density is derived much more easily via the energy averaging
method than in the equal-time representation. Proceeding to spinor
QED, we derive the covariant version of the equal-time equation
derived by Bialynicki-Birula et al. We show that it must be
supplemented by another self-adjoint equation to obtain a complete
description of the covariant spinor Wigner operator. After spinor
decomposition and energy average we study the classical limit of the
resulting three-dimensional kinetic equations. There are only two
independent spinor components in this limit, the mass density and the
spin density, and we derive also their covariant equations of motion.
We then show that the equal-time kinetic equation provides a complete
description only for constant external electromagnetic fields, but is
in general incomplete. It must be supplemented by additional
constraints which we derive explicitly from the covariant formulation.

\end{abstract}

\pagestyle{plain}
\pagenumbering{arabic}
\baselineskip 24pt plus 4pt minus 4pt
\vglue 0.2in

\def\boldgamma{{\mbox{\boldmath$\gamma$}}}
\def\half{{\textstyle{1\over 2}}}
\def\quarter{{\textstyle{1\over 4}}}

 \section{Introduction}

Theoretic evidence from lattice simulations \cite{Re1} of QCD for
a phase transition from a hadron gas to a quark-gluon plasma at high
temperature has prompted experimental efforts to create this new phase
of matter in the laboratory during the early stages of
ultrarelativistic heavy ion collisions \cite{Re2}. However, because of
the estimated very short life time of the collision zone, the highly
excited quark-gluon system may spend a considerable fraction of its
life in a non-thermalized, pre-equilibrium state. The dynamical tool
to treat dissipative processes in heavy ion collisions and the
approach to local thermal equilibrium is in principle non-equilibrium
quantum transport theory. Since QCD, the theory describing the
interactions between quarks and gluons, is a gauge theory, the kinetic
equations should be gauge covariant. A relativistic and gauge
covariant kinetic theory for quarks and gluons has been derived
\cite{Re3}, both in a classical framework \cite{Re4,Re5} and as a
quantum kinetic theory \cite{Re6,Re7} based on Wigner operators
defined in 8-dimensional phase space \cite{Re8}. Some preliminary
applications to the quark-gluon plasma, such as linear color response,
color correlations and collective plasma oscillations \cite{Re9,Re10}
have been discussed in this framework using a semiclassical expansion
of the quantum transport theory. Apart from its relativistic and
gauge covariance, an important aspect of the four-dimensional
approach to transport theory is that the complex kinetic equation can
be split up into a constraint and a transport equation, where the
former is a quantum extension of the classical mass-shell condition,
and the latter is a covariant generalization of the Vlasov-Boltzmann
equation. The complementarity of these two ingredients is essential
for a physical understanding of quantum kinetic theory \cite{henning}.

As a first step towards a full kinetic treatment of the quark-gluon
plasma, a detailed study of relativistic transport theory
for electrodynamics was performed over the years (see \cite{Re3,hakim}
for references). The absence of the complications arising from the
non-abelian character of QCD allows for a deeper insight into
the structure of the kinetic theory itself, and it is also easier to
find useful applications in this simpler case. Vasak, Gyulassy and
Elze \cite{Re11} (to be quoted as VGE) discussed the relativistic
quantum transport theory for spinor electrodynamics and gave a very
lucid account of the covariant spinor decomposition for the Wigner
operator of Dirac particles (see also \cite{hakim}).

In classical transport theory, all the physical currents are connected
with the distribution function $f$. The quantum mechanical analogue of
$f$ is the Wigner function, which is the ensemble expectation of the
Wigner operator. In the process of performing the ensemble average of
the kinetic equation for the Wigner operator, one encounters the
two-body correlation function $\langle F^{\mu\nu}\hat W \rangle$,
where $F^{\mu\nu}$ is the gauge field strength. In the general case, the
two-body correlation function again depends on higher order
correlation functions; this generates the so called BBGKY hierarchy
\cite{Re8}. A popular method to obtain a closed kinetic equation for
the Wigner function $W = \langle \hat W \rangle$ is to use the Hartree
approximation where the gauge field is considered as a mean field
$\bar F^{\mu\nu}$, leading to the replacement $\langle F^{\mu\nu} \hat
W \rangle = \bar F^{\mu\nu} \langle \hat W \rangle$. In this
approximation the BBGKY chain is truncated at the one body level, and
the kinetic equation for the Wigner function has the same form as that
for the Wigner operator. This version of the mean field approximation,
which treats the particle fields quantum mechanically, but uses the
classical approximation for the gauge fields, is widely used in
electrodynamic transport theory \cite{Re11,Re12,Re13}. It is
appropriate for strong but slowly varying electromagnetic fields.  An
interesting example is the Schwinger process \cite{Re14,Re15} for pair
production in external electromagnetic fields. Since this
approximation is sufficient for our purpose, we will also restrict
ourselves to the investigation of transport theory for particles
interacting with an external field.

In a recent paper Bialynicki-Birula, Gornicki and Rafelski \cite{Re12}
(to be quoted as BGR) proposed an equal-time transport equation for
spinor electrodynamics. Unlike the covariant theory where the Wigner
operator is defined as the four-dimensional Fourier transform of the
gauge covariant density matrix $\Phi_4(x,y)$ (see below),
 \begin{equation}
 \label{w4}
   \hat W_4(x,p) = \int d^4y\, e^{ip{\cdot}y} \, \Phi_4(x,y) \, ,
 \end{equation}
they introduced an equal-time correlation function $\Phi_3 (x,{\bf y})
= \Phi({\bf x},{\bf y},t)$ and considered its spatial Fourier
transform which depends only on three momentum coordinates:
 \begin{equation}
 \label{w3}
   \hat W_3(x,{\bf p}) = \int d^3y\,
   e^{-i{\bf p}{\cdot}{\bf y}} \, \Phi_3(x,{\bf y}) \, .
 \end{equation}
{}From this definition and the Dirac equation they obtained directly a
kinetic equation for $\hat W_3$. They showed that the Wigner function
$\langle \hat W_3(x,{\bf p}) \rangle$ is a direct analogue of the
classical distribution function $f({\bf x},{\bf p},t)$ and that it
provides a systematic way of studying the phase-space dynamics of QED
in the semiclassical limit; each spinor component of $\langle \hat W_3
\rangle$ corresponds to a definite physical distribution function. Of
course, $\hat W_3(x,{\bf p})$ is not manifestly Lorentz covariant.

Recently this equal-time method has been extended \cite{Re13} to
scalar electrodynamics in the Feshbach-Villars representation. The
ensuing applications of this so-called ``three-dimensional" transport
theory to the problem of nonperturbative pair creation in both spinor
\cite{Re12} and scalar \cite{Re16,Re17} QED indicated that some quantum
problems may be treated more easily in phase space in terms of the
Wigner operator than by conventional field theoretic methods. We will
show, however, that the published derivation of the equal-time
transport theory is incomplete. A complete treatment leads to
additional constraints on the Wigner function which for the specific
situations treated by previous authors actually simplifies its
structure; in general, however, it leads to complications whose
effects cannot be neglected.

We show in this paper that the three-dimensional transport theory can
also be derived by a different method which does not suffer from this
incompleteness. In particular, we discuss the relationship between the
covariant and three-dimensional approaches. From the comparison of
Eqs.(\ref{w4}) and (\ref{w3}) it is easily seen that there exists a
general connection between the four- and three-dimensional Wigner
operators:
 \begin{equation}
 \label{w34}
   \hat W_3(x,{\bf p}) = \int dp_0\ \hat W_4(x,p)\, .
 \end{equation}
This relation shows that the three-dimensional transport theory can
be obtained by an energy average of its covariant version. The main
formal difference between the energy averaging method used in this
paper and the direct equal-time derivation of BGR is that our approach
starts with an explicitly relativistic formulation. This covariant
formulation is shown to be complete, i.e. the equation of motion for
the Wigner operator is equivalent to the Dirac equation for the field
operators. This completeness is preserved during the reduction from
the four- to the three-dimensional version via energy averaging. The
covariant formulation followed by energy averaging also allows us to
obtain a three-dimensional transport equation for the scalar Wigner
operator directly from the Klein-Gordon equation, rather than via the
Feshbach-Villars $2\times 2$ matrix formulation.

We proceed as follows. In section 2, we discuss the transport
theory for spinless charged particles. We first set up the covariant
constraint and transport equations, which are the basis of the
reduction to the three-dimensional version. Next we consider the
semiclassical expansion in $\hbar$ and the classical limit. Then we
average the covariant equations over the energy to obtain the
three-dimensional equations. With these we investigate the pair
creation of charged scalar particles in a spatially homogeneous but
time-dependent electric field. This problem was studied before
in the Feshbach-Villars matrix representation \cite{Re17}, but our
procedure turns out to be much simpler because of the scalar nature of
our Wigner function. In Section 3, we study the kinetic theory for
Dirac fermions. We derive a complete set of two selfadjoint covariant
equations, one of which corresponds to the four-dimensional version of
the BGR equation. We give a full mapping of the spinor components from
the four- to the three-dimensional representation. The two selfadjoint
covariant transport equations can be combined into a single complex
equation, the VGE equation, which is then subjected to the energy
averaging procedure. We discuss the classical limit of the resulting
three-dimensional equations and compare it with the results
\cite{Re18} from the equal-time method. We then show how to get a
complete set of three-dimensional kinetic equations in the general,
fully quantum mechanical case; these contain as a subset the various
spinor components of the BGR equation, but also a set of additional
constraints. We discuss these results and summarize them in our
conclusions.

 \section{Scalar electrodynamics}
 \subsection{Relativistic covariant kinetic equations}

The abelian gauge theory of scalar particles with mass $m$ and charge
$e$ is defined through the Lagrangian density
 \begin{equation}
 \label{Sl}
   {\cal L} = (\partial_\mu-ieA_u)\phi^\dagger
              (\partial^\mu+ieA^\mu)\phi - m^2\phi^\dagger\phi
              - \quarter F_{\mu\nu} F^{\mu\nu} \ .
 \end{equation}
It leads to the Klein-Gordon equation for the scalar field operator
$\phi$,
 \begin{equation}
 \label{Klein}
   \left[ (\partial_\mu + ieA_\mu(x)) (\partial^\mu+ieA^\mu(x))
   + m^2 \right] \, \phi(x) = 0 \ ,
 \end{equation}
the adjoint equation for $\phi^\dagger$, and to Maxwell's equations
for the field strength tensor $F^{\mu\nu}$,
 \begin{equation}
 \label{Ma}
   \partial_\mu F^{\mu\nu}(x) = j^\nu (x)  \ ,
   \qquad\qquad
   \partial_\mu \tilde F^{\mu\nu}(x) = 0 \ ,
 \end{equation}
where $\tilde F^{\mu\nu}(x) = {1\over 2} \epsilon^{\mu\nu\rho\sigma}
F_{\rho\sigma}$ is the dual field tensor and $j^\nu(x)$ is the charge
current of the scalar particles.

The relativistically covariant Wigner operator is the Fourier
transform Eq.~(\ref{w4}) of the gauge covariant scalar field
correlation function
 \begin{equation}
 \label{Sd1}
   \Phi_4(x,y) = \phi(x)e^{-y\cdot{\cal D}^\dagger/2}\cdot
                        e^{y\cdot{\cal D}/2}\phi^\dagger(x) \ .
 \end{equation}
The covariant derivatives ${\cal D}_\mu(x)=\partial_\mu + ieA_\mu(x)$
and ${\cal D}^\dagger_\mu(x) = \partial_\mu - ieA_\mu(x)$ in this
expression ensure the gauge covariance (actually: gauge independence
in the case of a $U(1)$ gauge theory) of the density matrix.
Following \cite{Re6} and replacing the covariant derivatives here by a
line integral of the gauge field $A_\mu(x)$ along a straight line
between the space-time points $x-{y\over 2}$ and $x+{y\over 2}$, we
recover the more familiar form \cite{Re14}
 \begin{equation}
 \label{Sd2}
   \Phi_4(x,y) = \phi\left(x+\half y \right) \,
                 \exp\left[ ie\int^{1\over 2}_{-{1\over 2}}ds\,
                             A(x+sy){\cdot}y \right]
                 \phi^\dagger\left(x-\half y\right) \, .
 \end{equation}
The resulting scalar Wigner operator is selfadjoint:
 \begin{equation}
 \label{Sw1}
   \hat W^\dagger_4(x,p) = \hat W_4(x,p) \ .
 \end{equation}

Substituting Eq.~(\ref{Sd1}) into Eq.~(\ref{w4}) and integrating over
$y$ we see that the Wigner operator is just the density of particles
at space-time point $x$ with kinetic momentum $p$,
 \begin{equation}
 \label{Sw2}
   \hat W_4(x,p) = \phi(x) \, \delta^4 (p-\hat\pi(x)) \,
                   \phi^\dagger(x) \, ,
 \end{equation}
where $\hat\pi(x) = {i\over 2}({\cal D}_\mu(x) - {\cal D}_\mu^\dagger
(x))$ is the kinetic momentum operator. Eq.~(\ref{Sw2}) motivates a
statistical interpretation of the Wigner function as a generalized
phase-space density and generates simple relations between the Wigner
operator and all physical space-time density operators. For example,
the operators for the charge current density and the energy momentum
tensor are given by
 \begin{eqnarray}
 \label{CE}
   \hat j_\mu(x)
   &=& ie \Bigl( \phi^\dagger(x) \partial_\mu \phi(x)
                - (\partial_\mu \phi^\dagger(x)) \phi(x) \Bigr)
   \nonumber\\
   &=& e \int d^4p \, p_\mu \, \hat W_4(x,p) \, ,
   \nonumber\\
   \hat T_{\mu\nu}(x)
   &=& \half \Big( (\partial_\mu\phi^\dagger(x))
                        (\partial_\nu \phi(x))
                      - (\partial_\nu \phi^\dagger(x))
                        (\partial_\mu\phi(x))\Big)
   \nonumber\\
   &=& \int d^4p\, p_\mu p_\nu \, \hat W_4(x,p) \, .
 \end{eqnarray}

The equations of motion for the Wigner operator are a direct
consequence of the field equations (\ref{Klein}). We calculate the
second-order derivatives of the correlator $\Phi_4(x,y)$ with respect
to $x$ and $y$,
 \begin{eqnarray}
 \label{Sd3}
   \left( \half \partial_\mu^x +\partial_\mu^y \right)
   \left( \half \partial_x^\mu +\partial_y^\mu \right) \!\!\!\!\!\!\!\!
   && \Phi_4(x,y) = - m^2\Phi_4(x,y)
   \nonumber\\
   &&- 2ie\int^{1\over 2}_{-{1\over 2}}ds
       \left(\half+s\right) y_\nu F^{\nu\mu}(x+sy)
       \left(\half\partial_\mu^x + \partial_\mu^y\right)
       \, \Phi_4(x,y)
   \nonumber\\
   && -(ie)^2 \left[\int^{1\over 2}_{-{1\over 2}}ds
              \left( \half + s \right)
              y{\cdot}F(x+sy)\right]^2
              \! \Phi_4(x,y)
   \nonumber\\
   && + ie\int ^{1\over 2}_{-{1\over 2}}ds
        \left(\half+s\right)^2
        y{\cdot}j(x+sy)\, \Phi_4(x,y) \, ,
 \end{eqnarray}
where we have employed the Klein-Gordon and Maxwell equations in the
first and last terms on the right-hand side. Eq.~(\ref{Sd3}) can be
rewritten in a compact way,
 \begin{equation}
 \label{Sd4}
   \left[ \quarter D^\mu D_\mu - \Pi^\mu\Pi_\mu + m^2
         - i\Pi^\mu D_\mu\right]\,\Phi_4(x,y) = 0\, ,
 \end{equation}
where we defined the two Lorentz covariant operators $D$ and $\Pi$ by
 \begin{eqnarray}
 \label{DP1}
   D^\mu(x,y)
   &=& \partial^\mu_x + ie\int^{1\over 2}_{-{1\over 2}}ds
       \, y_\nu F^{\nu\mu}(x+sy) \, ,
   \nonumber\\
   \Pi^\mu(x,y)
   &=& i\left(\partial^\mu_y + ie\int^{1\over 2}_{-{1\over 2}}
       ds\, s\, y_\nu F^{\nu\mu}(x+sy)\right) \, .
 \end{eqnarray}

Performing the Fourier transform with respect to $y$, we obtain the
following exact quadratic kinetic equation for the scalar Wigner
operator:
 \begin{equation}
 \label{Sw3}
   \left[ \quarter \hbar^2 D^\mu D_\mu - \Pi^\mu \Pi_\mu + m^2
         -i\hbar\Pi^\mu D_\mu \right]\, \hat W_4(x,p) = 0 \, .
 \end{equation}
The operators $D$ and $\Pi$ in this equation are now defined in phase
space and can be obtained from Eqs.~(\ref{DP1}) by the replacements
$y^\mu \to -i\partial_p^\mu$ and $\partial_y^\mu \to -ip^\mu$:
 \begin{eqnarray}
 \label{DP2}
    D_\mu(x,p)
    &=& \partial_\mu - e\int ^{1\over 2}_{-{1\over 2}} ds\,
        F_{\mu\nu} (x-i\hbar s\partial_p)\, \partial_p^\nu \, ,
    \nonumber \\
    \Pi_\mu(x,p)
    &=& p_\mu - ie\hbar \int ^{1\over 2}_{-{1\over 2}} ds\, s\,
        F_{\mu\nu}(x-i\hbar s\partial_p) \, \partial_p^\nu \, .
 \end{eqnarray}
In Eqs.~(\ref{Sw3}) and (\ref{DP2}) we reinstated the
$\hbar$-dependence explicitly in order to be able to discuss the
semiclassical expansion in the next subsection. (The speed
of light $c$ is still omitted.) Obviously, the operators $D_\mu$ and
$\Pi_\mu$ are gauge covariant extensions of the partial derivative
$\partial_\mu$ and the momentum $p_\mu$, respectively. They both are
self-adjoint: namely, $D_\mu^\dagger(x,p)=D_\mu(x,p)$ and
$\Pi_\mu^\dagger(x,p) = \Pi_\mu(x,p)$.

Since the scalar Wigner operator is self-adjoint, too, the real and
imaginary parts of the complex equation (\ref{Sw3}) have to vanish
separately:
 \begin{eqnarray}
 \label{Sc}
   &&\left(\quarter \hbar^2 D^\mu D_\mu - \Pi^\mu\Pi_\mu + m^2\right)
     \hat W_4(x,p) = 0 \,  ,
 \\
 \label{St}
   &&\hbar\Pi^\mu D_\mu \hat W_4(x,p) = 0 \, .
 \end{eqnarray}
The first equation is usually called constraint equation since
it is obviously a generalization of the classical mass-shell condition
$p^2-m^2=0$. The second equation has the typical form of a transport
equation and is the generalization of the classical Vlasov equation
for charged particles with abelian interactions $p^\mu (\partial_\mu -
e F_{\mu\nu}(x) \partial_p^\nu) f(x,p) = 0$, where $f(x,p)$ is the
classical distribution function. These two equations together give a
complete description for the Wigner operator; they are equivalent
to the original field equations of motion.

 \subsection{Semiclassical expansion}

In order to better understand the structure of the constraint and
transport equations, we consider their semiclassical expansion in
$\hbar$ and their classical limit. The calculation of quantum
corrections to this limit can then be performed in a systematic way.

As stated in the Introduction, we will in this paper consider the
electromagnetic field as a classical (mean) field. In this
approximation the equation of motion for the Wigner {\em operator} and
its ensemble expectation value, the Wigner {\em function}, become
formally identical. Therefore we will now work with the equations of
motion for the Wigner {\em function} by leaving off the hats over the
corresponding operators.

The field strength $F^{\mu\nu}$ in Eqs.~(\ref{DP2}), which has to be
evaluated at the shifted argument $x-i\hbar s\partial_p$, is defined in
terms of its Taylor expansion around $x$ and can be expressed in terms
of the so-called ``triangle operator" $\triangle =
\partial_p{\cdot}\partial_x$ as
 \begin{equation}
 \label{Fs}
   F_{\mu\nu}(x-is\hbar\partial_p) =
      e^{is\hbar \triangle } F_{\mu\nu}(x)\, .
 \end{equation}
The $s$-integration can then be done, and expanding the result in
powers of $\hbar$ we obtain
 \begin{eqnarray}
 \label{DP3}
   D_\mu(x,p)
   &=& \partial_\mu - e{\sin(\hbar \triangle /2)
                        \over \hbar \triangle/2}
       F_{\mu\nu}(x)\, \partial_p^\nu
 \nonumber\\
   &=& \Big(\partial_\mu -eF_{\mu\nu}(x)\,\partial_p^\nu \Big)
      +{e\over 24} \hbar^2\triangle^2F_{\mu\nu}(x)\,\partial_p^\nu
      +\cdots \, ,
 \nonumber\\
   \Pi_\mu(x,p)
   &=& p_\mu +{e\hbar\over 2}\left({\cos(\hbar\triangle/2)
                                    \over \hbar \triangle/2}
                 -{\sin(\hbar\triangle/2)
                   \over (\hbar\triangle/2)^2}\right)
       F_{\mu\nu}(x)\, \partial_p^\nu
 \nonumber\\
   &=& p_\mu - {e\over 12}\hbar^2\triangle F_{\mu\nu}(x)\,\partial_p^\nu
                 +\cdots \, ,
 \end{eqnarray}
where the dots indicate corrections from higher orders of $\hbar$ or,
equivalently, of the derivative operator $\triangle$.

Expanding the Wigner function similarly in powers of $\hbar$,
 \begin{equation}
 \label{Sw4}
   W_4(x,p) = W_4^{(0)}(x,p) + \hbar W_4^{(1)}(x,p) +
                               \hbar^2 W_4^{(2)}(x,p) + \cdots \, ,
 \end{equation}
and inserting these expressions into the constraint and transport
equations (\ref{Sc}) and (\ref{St}), we obtain their semiclassical
expansion. In the zeroth order, there is no information from the
transport equation, and the constraint equation reduces to the
classical mass-shell condition,
 \begin{equation}
 \label{Ms1}
   (p^2-m^2)\, W_4^{(0)}(x,p) = 0 \, .
 \end{equation}
This equation has two elementary solutions corresponding to positive
and negative energies, and we can write
 \begin{equation}
 \label{Sw5}
    W_4^{(0)}(x,p) = W_4^{+(0)}(x, {\bf p})\,\delta (p_0-E_p)
       + W_4^{-(0)}(x, {\bf p})\,\delta (p_0+E_p) \, ,
 \end{equation}
where $E_p = + \sqrt{{\bf p}^2+m^2}$ is the classical on-shell energy.

To next order in $\hbar$, the transport equation (\ref{St}) begins to
contribute. It yields the Vlasov equation
 \begin{equation}
 \label{Ve1}
   p^\mu\Big(\partial_\mu-eF_{\mu\nu}(x)\partial_p^\nu\Big)
   W_4^{(0)}(x,p) = 0  \, ,
 \end{equation}
which must be evaluated with the ansatz (\ref{Sw5}). The constraint
equation yields at order $\hbar$
 \begin{equation}
 \label{Ms1a}
   (p^2-m^2)\, W_4^{(1)}(x,p) = 0 \, ,
 \end{equation}
with the solution
 \begin{equation}
 \label{Sw6}
    W_4^{(1)}(x,p) = W_4^{+(1)}(x, {\bf p})\,\delta (p_0-E_p)
       + W_4^{-(1)}(x, {\bf p})\,\delta (p_0+E_p) \, .
 \end{equation}

At order $\hbar^2$, Eq.~(\ref{Sc}) becomes
 \begin{equation}
 \label{Sw7}
  \!\!\!\!\!\!\!\!\!\!\!\!
  (p^2-m^2)\, W_4^{(2)}(x,p) =
  \left[\quarter\Big(\partial_x -eF(x){\cdot}\partial_p\Big)^2
       +{\textstyle{e\over 12}}
         \Big(p^\mu\triangle F_{\mu\nu}\,\partial_p^\nu
       +j(x){\cdot}\partial_p^\nu\Big)\right]
    W_4^{(0)}(x,p) \, ,
 \end{equation}
where $j_\nu(x) = e \int d^4p\, p_\nu\, W(x,p)$ is the average charge
current of the scalar particles in the ensemble which generates the
mean electromagnetic field via Maxwell's equation (\ref{Ma}).
Eq.~(\ref{St}) yields at order $\hbar^2$
 \begin{equation}
 \label{Ve1a}
    p^\mu\Big(\partial_\mu-eF_{\mu\nu}(x)\partial_p^\nu\Big)
      W_4^{(1)}(x,p) = 0  \,  .
 \end{equation}
The right-hand side of (\ref{Sw7}) represents off-shell corrections to
the mass-shell condition due to quantum corrections. Eq.~(\ref{Ve1a})
is the Vlasov equation for the first order Wigner operator. Quantum
corrections to the Vlasov equation for $W_4^{(2)}$ will arise only at
order $\hbar^3$. The fact that quantum corrections affect the Wigner
function only at second order in $\hbar$ is a specific feature of
scalar particles. In spinor QED quantum effects arise in first order
of $\hbar$ due to spin interactions.

We now carry out the $p_0$-average of the Vlasov equations
(\ref{Ve1},\ref{Ve1a}). This is facilitated by their simple form
(\ref{Sw5},\ref{Sw6}) resulting from the mass-shell conditions
(\ref{Ms1},\ref{Ms1a}). For the corresponding three-dimensional
positive and negative energy Wigner functions $W_3^\pm$ (see
Eq.~(\ref{w34})) we obtain
 \begin{equation}
 \label{Ve2}
   \partial_t W_3^{\pm(i)}(x,{\bf p})
   \pm ({\bf v}{\cdot}{\bf \nabla}) W_3^{\pm(i)}(x,{\bf p})
   + e \Big({\bf E}(x) \pm {\bf v}\times
             {\bf B}(x)\Big){\cdot}{\bf \nabla}_p
   W_3^{\pm(i)}(x,{\bf p}) = 0 \, ,
 \end{equation}
where ${\bf v}={\bf p}/E_p$, $i=0,1$, and ${\bf E}$ and ${\bf B}$ are
the electric and magnetic field components of
$F^{\mu\nu}$. We now introduce classical particle and
antiparticle distribution functions, $f$ and $\bar f$, via
 \begin{eqnarray}
 \label{Ve3}
  f^{(i)}(x,{\bf p}) &=& W_3^{+(i)}(x,{\bf p}) \, ,
  \nonumber\\
  \bar f^{(i)}(x,{\bf p}) &=& W_3^{-(i)}(x,{-\bf p}) \, .
 \end{eqnarray}
They satisfy the well-known three-dimensional Vlasov transport
equations for scalar particles and antiparticles moving in an external
electromagnetic field:
 \begin{eqnarray}
 \label{Ve4}
   \partial_t f^{(i)}(x,{\bf p})
   + ({\bf v}{\cdot}{\bf \nabla})f^{(i)}(x,{\bf p})
   + e\Big({\bf E}(x)+{\bf v}\times {\bf B}(x)
           \Big){\cdot}{\bf \nabla}_p f^{(i)}(x,{\bf p}) = 0\, ,
 \nonumber \\
   \partial_t \bar f^{(i)}(x,{\bf p})
   + ({\bf v}{\cdot}{\bf \nabla})\bar f^{(i)}(x,{\bf p})
   - e\Big({\bf E}(x)+{\bf v}\times {\bf B}(x)
           \Big){\cdot}{\bf \nabla}_p \bar f^{(i)}(x,{\bf p}) = 0
   \, .
 \end{eqnarray}

 \subsection{Pair production in transport theory}

In this subsection, we give an exact non-perturbative solution of the
scalar transport theory in a spatially constant external electric
field. The solution describes pair production due to vacuum
excitation. In electrodynamics, it was first studied many years ago by
Schwinger \cite{Re14}, who connected the probability of pair creation
with the imaginary part of the effective action in QED. Recently it
was reinvestigated as an application of the equal-time transport
theory. For scalar electrodynamics in the Feshbach-Villars
representation, Best and Eisenberg \cite{Re17} obtained from the
kinetic theory the same result as Popov did previously in \cite{Re19}
with field operator techniques. Here we consider the pair creation of
scalar particles directly in the Klein-Gordon representation of the
kinetic theory, to show how the energy averaging method works
in a non-perturbative case and to discuss its difference from the
Feshbach-Villars based equal-time method.

 \subsubsection{Initial value problem}

The determination of any solution of the differential equations
(\ref{Sc}) and (\ref{St}) needs initial conditions. For the pair
creation problem we should search for a free vacuum solution as the
initial condition. For ${\bf E}={\bf B}=0$, the energy average of the
full constraint and transport equations result in the following
three-dimensional expressions:
 \begin{eqnarray}
 \label{Ec}
    \Bigl(\partial_t^2 -{\bf \nabla}^2 +4E_p^2\Bigr) W_3(x,{\bf p})
     &=& 4\, \varepsilon(x,{\bf p}) \, ,
 \nonumber\\
    \partial_t \, \rho(x,{\bf p}) - {\bf \nabla}{\cdot}{\bf j}(x,{\bf p})
    &=& 0 \, .
 \end{eqnarray}
Here we defined the phase-space densities of electric charge
$\rho(x,{\bf p})$, electric current ${\bf j}(x,{\bf p})$, and energy
$\varepsilon(x,{\bf p})$ by
 \begin{eqnarray}
 \label{Ecc}
    \rho(x,{\bf p})
    &=& \int dp_0\, j_0(x,p) = e\int dp_0\, p_0 \, W_4(x,p) \, ,
 \nonumber\\
    {\bf j}(x,{\bf p})
    &=& \int dp_0 \, {\bf j}(x,p) = e\, {\bf p} \, W_3(x,{\bf p}) \, ,
 \nonumber\\
    \varepsilon (x,{\bf p})
    &=& \int dp_0\, T_{00}(x,p) = \int dp_0 \, p_0^2 \, W_4(x,p) \, .
 \end{eqnarray}

The simplest homogeneous solution of Eq.~(\ref{Ec}) is
 \begin{eqnarray}
 \label{Va}
    W_3(x,{\bf p})  &=& {1\over E_p} \, ,
 \nonumber \\
    \rho(x,{\bf p}) &=& 0 \, ,
 \nonumber \\
    {\bf j}(x,{\bf p}) &=& e{\bf v}=e{{\bf p}\over E_p}\, ,
 \nonumber\\
    \varepsilon(x,{\bf p}) &=& E_p \, .
 \end{eqnarray}

Since purely magnetic fields do not produce pairs, we can restrict our
attention for the Schwinger pair creation mechanism to electric
fields. We assume for simplicity a spatially homogeneous but
time-dependent electric field. The spatial homogeneity of the external
fields and the initial condition (\ref{Va}) allow to reduce the
constraint and transport equations to ordinary differential equations
in time. The three-dimensional kinetic equations are obtained through
the energy average as
 \begin{eqnarray}
 \label{P1}
    \Bigl(D_t^2 + 4E_p^2
          + {e\over 3}\,(\partial_t {\bf E}){\cdot}{\bf \nabla}_p
    \Bigr)\, W_3(t,{\bf p})
    &=& 4\, \varepsilon (t,{\bf p}) \, ,
 \nonumber \\
    D_t \rho(t,{\bf p}) &=& 0 \, ,
\end{eqnarray}
where
 \begin{equation}
 \label{P2}
   D_t = \partial_t+e{\bf E}\cdot{\bf \nabla}_p \, .
 \end{equation}

To obtain a closed equation of motion for the Wigner function
$W_3(t,{\bf p})$, we eliminate the energy density $\varepsilon$ from the
first equation of (\ref{P1}). To this end we multiply the transport
equation (\ref{St}) by $p_0$ from the left and then integrate it with
respect to $p_0$:
 \begin{equation}
 \label{P3}
   D_t \varepsilon (t,{\bf p}) =
   \left({e\over 12}(\partial_t {\bf E}){\cdot}{\bf \nabla}_p D_t
       + {e\over 12}(\partial^2_t{\bf E}){\cdot}{\bf \nabla}_p
       + e {\bf p}{\cdot}{\bf E}\right)
   W_3(t,{\bf p}) \, .
 \end{equation}
We then apply the operator $D_t$ on the first equation (\ref{P1}) and
combine it with Eq.~(\ref{P3}) to eliminate $\varepsilon$. We obtain
 \begin{equation}
 \label{P4}
   \Bigl( D_t^3 + 4E_p^2D_t + 4e{\bf E}{\cdot}{\bf p} \Bigr)
   W_3(t,{\bf p})=0 \ ,
 \end{equation}
which, together with the initial condition
 \begin{equation}
 \label{P4a}
   W_3(t=-\infty)={1\over E_p} \, ,
 \end{equation}
can be solved as an initial value problem for the Wigner function.
We assume that the external electric field is switched on
adiabatically in the far past and switched off adiabatically in the
far future, ${\bf E}(t=-\infty)={\bf E}(t=\infty)=0$.

The momentum derivative hidden in the operator $D_t$ complicates the
solution of Eq.~(\ref{P4}). We follow \cite{Re12} and use the
well-known method of characteristics in order to separate the momentum
derivative and obtain an ordinary differential equation in time. One
introduces a test Wigner function \cite{Re12} $\omega_3$ through
 \begin{equation}
 \label{P5}
    W_3(t,{\bf p}) = \int d^3{\bf p}_0\, \omega_3(t,{\bf p}_0) \,
    \delta^{(3)}\bigl({\bf p}-{\bf p}(t,{\bf p}_0)\bigr) \, ,
 \end{equation}
where the function ${\bf p}(t,{\bf p}_0)$ is a solution of the
classical equation of motion for a particle with initial momentum
${\bf p}_0$ in an external electric field:
 \begin{equation}
 \label{P6}
   {d{\bf p}(t,{\bf p}_0)\over dt}=e{\bf E}(t) \, .
\end{equation}
Its explicit form reads
 \begin{equation}
 \label{P7}
   {\bf p}(t,{\bf p}_0)={\bf p}_0+e\int_{-\infty}^t dt'\, {\bf E}(t') \, .
 \end{equation}
Substituting (\ref{P5}) into (\ref{P4}), the momentum derivative is
absorbed into the classical motion, and the partial differential equation
is converted into
 \begin{equation}
 \label{P8}
   \int d^3{\bf p}_0\, \left[\Bigl(\partial^3_t + 4E_p^2\partial_t
      + 4e{\bf E}{\cdot}{\bf p}\Bigr) \omega_3 (t,{\bf p}_0) \right]\,
      \delta^{(3)}\Bigl({\bf p}-{\bf p}(t,{\bf p}_0)\Bigr)=0 \, .
 \end{equation}
This is solved if $\omega_3(t,{\bf p}_0)$ satisfies the ordinary
differential equation
 \begin{equation}
 \label{P9}
   \Bigl(\partial^3_t + 4E_p^2(t,{\bf p}_0)\,\partial_t
                     + 4e{\bf E}{\cdot}{\bf p}(t,{\bf p}_0)\Bigr)
   \omega_3(t,{\bf p}_0)=0
 \end{equation}
with the initial condition
 \begin{equation}
 \label{P9a}
   \omega_3(t=-\infty,{\bf p}_0)={1\over E_p} \, ,
 \end{equation}
where ${\bf p}(t,{\bf p}_0)$ satisfies Eq.~(\ref{P7}) and $E_p(t,{\bf
p}_0)$ is the correspondimg on-shell energy.

Please note the time dependence of the momentum ${\bf p}$ and the
particle energy $E_p$ arising from the classical equation of motion:
when inserting the solution of (\ref{P9}) into (\ref{P5}) in order to
construct the Wigner function $W_3$ from the test function $\omega_3$,
all the classical time dependence of $W_3$ resides in these functions
${\bf p}(t,{\bf p}_0)$, $E_p(t,{\bf p}_0)$, while the additional time
dependence from quantum effects is described by the differential
equation (\ref{P9}) for $\omega_3$.

For later it will be useful to similarly introduce a test charge
density $\varrho (t,{\bf p}_0)$, a test charge current ${\bf \cal
J}(t,{\bf p}_0)$, and a test energy density ${\bf \cal E}(t,{\bf
p}_0)$; for example,
 \begin{equation}
 \label{P90}
   \rho(t,{\bf p})=\int d^3{\bf p}_0\, \varrho (t,{\bf p}_0)\,
   \delta^{(3)}\Bigl({\bf p}-{\bf p}(t,{\bf p}_0)\Bigr) \, .
 \end{equation}
They satisfy for arbitrary initial momentum ${\bf p}_0$ the following
equations:
 \begin{eqnarray}
 \label{P10}
   \partial_t\, \varrho(t,{\bf p}_0)
   &=& 0 \, ,
 \nonumber \\
   \partial_t\, {\bf \cal E}(t,{\bf p}_0)
   &=& {\bf E}\cdot {\bf \cal J}(t,{\bf p}_0) \, ,
\end{eqnarray}
which follow from Eqs.(\ref{P1}) and (\ref{P3}), respectively. Their
physical meaning is evident: The first equation shows that the net
charge density always vanishes due to the homogeneous initial
conditions (\ref{Va}). The second one is just Poynting's theorem of
energy-momentum conservation.

 \subsubsection{Pair density}

As pointed out in \cite{Re19}, the problem of pair creation in a
homogeneous, but time-dependent electric field can be mapped onto the
quantum mechanics problem of an oscillator with variable frequency.
This will be exploited in the following treatment. We introduce an
auxiliary function $\zeta(t)$ via
 \begin{equation}
 \label{O1}
   \omega_3(t) = |\zeta(t)|^2 \, ,
 \end{equation}
in terms of which the pair production problem (\ref{P9},\ref{P9a})) is
reduced to the solution of a quantum oscillator problem:
 \begin{equation}
 \label{O2}
   \partial^2_t \zeta + E_p^2(t) \, \zeta=0\, ,\qquad
   |\zeta(t=-\infty)|= 1/\sqrt{E_p} \, .
 \end{equation}
Since we are interested in the total pair production yield at time
$t\rightarrow \infty$, it is sufficient to study the asymptotic
solutions of this equation. Due to its similarity with the
time-dependent barrier-potential problem \cite{Re20} in
non-relativistic quantum mechanics, the asymptotic solutions can be
easily written down in WKB approximation:
 \begin{eqnarray}
 \label{O4}
    \zeta(t\to -\infty)
    &=& e^{-iE_p^{-\infty}t} / \sqrt{E_p^{-\infty}} \, ,
 \nonumber \\
    \zeta(t\to + \infty)
    &=& C_1e^{-iE_p^\infty t}+C_2e^{iE_p^\infty t} \, .
\end{eqnarray}
Here we have already taken into account the initial condition for
$\zeta$, and $E_p^{-\infty}$ and $E_p^\infty$ are the asymptotic
particle energies $E_p^{-\infty}=\sqrt{m^2+{\bf p}_0^2}\,$ and
$E_p^\infty = \sqrt{m^2+({\bf p}(t=\infty,{\bf p}_0))^2}$. The
condition for applicability of the WKB approximation \cite{Re21},
 \begin{equation}
 \label{O3}
   \partial_t E_p = {e{\bf p}{\cdot}{\bf E}\over E_p} \ll E_p^2 \, ,
 \end{equation}
is satisfied in the limit $t\to \pm \infty$ since the electric field
vanishes in this limit.

{}From these asymptotic solutions we see that the test Wigner function
in the limit $t\to +\infty$ contains both oscillating and
non-oscillating parts:
 \begin{equation}
 \label{O5}
    \omega_3(t\to \infty) = \vert \zeta(t\to \infty) \vert^2
    = (|C_1|^2+|C_2|^2)+(C_1C_2^*e^{-2iE_p^\infty t} + {\rm c.c.}) \, .
 \end{equation}
The created pairs are separated and accelerated by the electric field
in opposite directions, thereby generating a current. Writing this
current as ${\bf \cal J}(t) = e{\bf v(t)} n(t)$, where $n(t)$ is the
total particle density (positive plus negative particles), and
comparing with its expression in terms of the test Wigner function,
namely, ${\bf \cal J}(t) = e{\bf p}(t)\, \omega_3(t)$, we find
 \begin{equation}
 \label{P11}
    n(t) = E_p(t)\, \omega_3(t) \, .
 \end{equation}
Taking the time average of Eq.~(\ref{O5}) which removes the rapidly
oscillating parts, we thus find for the asymptotic value of the total
particle density
 \begin{equation}
 \label{P12}
    n(t=\infty)=E_p^\infty \Bigl( |C_1|^2 + |C_2|^2 \Bigr) \, .
 \end{equation}
Due to the conservation law
 \begin{equation}
 \label{P13}
   \partial_t\Big[(\partial_t \zeta) \zeta^* -
                  (\partial_t \zeta^*) \zeta \Big] = 0 \, ,
 \end{equation}
which is easily derived from (\ref{O2}), the expression in the square
bracket is the same at $t=+\infty$ and $t=-\infty$ and can thus be
evaluated with the initial conditions (\ref{O4}). We obtain
 \begin{equation}
 \label{P14}
   |C_1|^2 - |C_2|^2 = {1\over E_p^\infty} \, .
\end{equation}
Thus the oscillator is fully characterized by the ratio
 \begin{equation}
 \label{P15}
    r = {|C_2|^2 \over |C_1|^2} \, ,
 \end{equation}
which can be interpreted (and calculated) as the transmission
coefficient of the non-relativistic barrier-potential problem
\cite{Re20}. In terms of this ratio, the asymptotic particle density
is given as
 \begin{equation}
 \label{P16}
   n(t=\infty) = {1+r\over 1-r} = 1+2{r\over 1-r} \, .
 \end{equation}
The first term on the right-hand side is a vacuum contribution and
stems from the Dirac sea of charged particles; it can not be measured
and must be removed by renormalization \cite{Re17}. The second term
arises from the pair creation. The factor of $2$ reminds us that both
particles and antiparticles contribute to the pair current. The
asymptotic pair density in phase-space is thus finally obtained as
 \begin{equation}
 \label{P17}
   n_{pair}(t=\infty) = {r \over 1-r} \, ,
 \end{equation}
which is in full agreement with the results derived both from field
theory \cite{Re19} and from the transport theory in Feshbach-Villars
representation \cite{Re17}.

Even though the final results of the two approaches to scalar
transport theory, the equal-time and the energy averaging methods, are
the same, we would like to point out one essential difference between
these two procedures: The equal-time formulation has to rely on the
Feshbach-Villars representation, since it requires field equations
which contain only first-order time derivatives. As a result, the
Wigner operator in the equal-time approach is a $2\times 2$ matrix
and first must be decomposed into its ``spinor'' components, in an
analogous way as required for spinor QED which will be discussed in
the following section. This matrix structure leads to a set of coupled
kinetic equations for the ``spinor'' components of $W_3$. The
energy averaging method works directly with the covariant Klein-Gordon
equation. This is a scalar equation, and no such complication arises.

 \section{Spinor electrodynamics}
 \subsection{Covariant version of the BGR equation}

We now turn to the investigation of the transport theory for
spin-${1\over 2}$ particles interacting with an
electromagnetic field. We start from the lagrangian density
 \begin{equation}
 \label{D1}
   {\cal L} = \bar\psi\Big(i\gamma^\mu(\partial_\mu+ieA_\mu)-m\Big)\psi
            - \quarter F^{\mu\nu}F_{\mu\nu} \, ,
 \end{equation}
which gives rise to the Dirac equations for the complex fields $\psi$
and $\bar\psi$,
 \begin{eqnarray}
 \label{D2}
   i\gamma^\mu(\partial_\mu+ieA_\mu)\psi
   &=& m\psi \, ,
 \nonumber\\
   i(\partial_\mu-ieA_\mu)\bar\psi\gamma^\mu
   &=& - m\bar\psi \, .
 \end{eqnarray}
The spinor Wigner operator is the four-dimensional Fourier transform
of the gauge invariant density matrix
 \begin{equation}
 \label{D3}
   \Phi_4(x,y)= \psi\left(x+\half y\right)
      \exp\left[ ie\int^{1\over 2}_{-{1\over 2}}ds\,
                 A(x+sy){\cdot}y \right]
      \bar\psi\left(x-\half y\right) \, .
\end{equation}
Unlike scalar electrodynamics, $\Phi_4$ is now a $4\times 4$ matrix in
spin space, and the Wigner operator $\hat W$ is no longer
self-adjoint. It behaves under hermitian conjugation like an
ordinary $\gamma$-matrix,
 \begin{equation}
 \label{D4}
   \hat W_4^\dagger(x,p)=\gamma_0\hat W_4(x,p)\gamma_0 \, .
 \end{equation}

The evolution equation of the covariant Wigner operator again follows
from the field equations of motion. Calculating the first-order
derivatives of the density matrix and using the Dirac equations in a
similar way as in the scalar case, we find a pair of equations for
$\Phi_4$ in terms of the operators $D$ and $\Pi$,
 \begin{eqnarray}
 \label{D5}
    \left( \half D_\mu(x,y) - i\Pi_\mu(x,y) \right)
         \gamma^0 \gamma^\mu \Phi_4(x,y)
    &=& - i m \gamma^0 \Phi_4(x,y) \, ,
 \nonumber\\
    \left( \half D_\mu(x,y) + i\Pi_\mu(x,y) \right)
         \Phi_4(x,y) \gamma^\mu \gamma^0
    &=& i m \Phi_4(x,y) \gamma^0 \, .
 \end{eqnarray}
The $\gamma_0$-matrices have been included in order to facilitate
comparison with the three-dimensional equal-time approach of
BGR \cite{Re12}.

Multiplying (\ref{D5}) by another $\gamma_0$-matrix from the right and
taking the Wigner transform, we derive the following kinetic equations
for the Wigner operator:
 \begin{eqnarray}
 \label{D6}
    &&\left( \half \hbar D_\mu(x,p) - i\Pi_\mu(x,p)\right)
            \gamma^0 \gamma^\mu \hat W_4(x,p) \gamma^0
        = - i m \gamma^0 \hat W_4(x,p)\gamma^0 \, ,
 \\
 \label{D61}
   &&\left( \half \hbar D_\mu(x,p) + i\Pi_\mu(x,p)\right)
           \hat W_4(x,p) \gamma^\mu
    = i m \hat W_4(x,p) \, .
 \end{eqnarray}
With an eye on the semiclassical expansion below, we have again
displayed the $\hbar$-depen\-dence explicitly. We note that these two
equations of motion in phase space, like the two Dirac equations
(\ref{D2}) in coordinate space, are adjoints of each other. Therefore,
either one of the Eqs.~(\ref{D6},\ref{D61}) provides a complete
description of the Wigner operator. By adding and subtracting these
two equations, respectively, we can get two different, now
self-adjoint equations:
 \begin{eqnarray}
 \label{D7}
    \half \hbar D_\mu \{ \gamma^0 \gamma^\mu,\ \hat W_4\gamma^0\}
    - i \Pi_\mu [\gamma^0\gamma^\mu, \ \hat W_4\gamma^0]
    &=& -i m [\gamma^0, \ \hat W_4\gamma^0] \, ,
 \\
 \label{D8}
    \half \hbar D_\mu [\gamma^0 \gamma^\mu, \ \hat W_4 \gamma^0]
    - i \Pi_\mu \{\gamma^0\gamma^\mu, \ \hat W_4\gamma^0 \}
    &=& - i m \{\gamma, \ \hat W_4\gamma^0 \} \, .
 \end{eqnarray}
Obviously (\ref{D7}) and (\ref{D8}) are symmetric under an exchange of
commutators and anticommutators. Their self-adjoint properties are
easily proven with the aid of Eq.~(\ref{D4}). Since these two
self-adjoint equations are equivalent to either of the equations
(\ref{D6},\ref{D61}), we have now three choices for a full description
of the Wigner operator: Eq.~(\ref{D6}), Eq.(\ref{D61}), or
Eq.(\ref{D7}) together with Eq.(\ref{D8}). Eqs.(\ref{D7}) and
(\ref{D8}) do not separately provide a complete description.

We would like to note another important feature of Eq.(\ref{D7}): The
commutator in the second term of the left-hand side leads to the
automatic disappearance of $p_0$ contained in the operator
$\Pi_0(x,p)$. Since the only remaining $p_0$-dependence is in $\hat
W_4$, this renders the energy average very simple. The result of an
integration over $p_0$ is just the BGR equation from the equal-time
formulation,
 \begin{equation}
 \label{D9}
    \!\!\!\!\!\!\!\!\!\!\!
    \hbar D_t \hat W_3(x,{\bf p}) = - \half \hbar
    {\bf D}{\cdot}\{ \gamma^0 \boldgamma, \ \hat W_3(x,{\bf p}) \}
    - i {\bf \Pi}{\cdot}[ \gamma^0 \boldgamma, \ \hat W_3(x,{\bf p})]
    - i m [\gamma^0, \ \hat W_3(x,{\bf p})] \, ,
 \end{equation}
with
 \begin{eqnarray}
 \label{D10}
    D_t(x,{\bf p})
    &=& \partial_t + e \int^{1\over 2}_{-{1\over 2}} ds\,
        {\bf E}({\bf x}+is\hbar{\bf \nabla}_p, t){\cdot}{\bf \nabla}_p
    \, ,
 \nonumber \\
    {\bf D}(x,{\bf p})
    &=& {\bf \nabla} + e \int^{1\over 2}_{-{1\over 2}}ds\,
    {\bf B}({\bf x}+is\hbar{\bf \nabla}_p, t) \times {\bf \nabla}_p
    \, ,
 \nonumber \\
    {\bf \Pi}(x,{\bf p})
    &=& {\bf p}-i e \hbar \int^{1\over 2}_{-{1\over 2}}ds\, s\,
    {\bf B}({\bf x} + is\hbar{\bf \nabla}_p, t) \times {\bf \nabla}_p
    \, ,
 \end{eqnarray}
where we have employed the BGR definition of the equal-time
Wigner operator \cite{Re12}:
 \begin{eqnarray}
 \label{D11}
    && \!\!\!\!\!\!\!\!\!\!
       \hat W_3(x,{\bf p}) = \int dp_0 \, \hat W_4(x,p)\, \gamma_0
 \\
    && = \int d^3{\bf y}\, e^{-i{\bf p}\cdot{\bf y}}\,
        \psi\left({\bf x}+\half{\bf y},t\right)\,
        \exp\left[ - i e \int^{1\over 2}_{-{1\over 2}}ds\,
                   {\bf A}({\bf x}+s{\bf y},t)\cdot{\bf p} \right]
        \psi^\dagger\left({\bf x}-\half{\bf y},t\right) \, .
 \nonumber
 \end{eqnarray}
Clearly equation (\ref{D7}) is the covariant version of the BGR
equation. However, as we have stressed, (\ref{D7}) is not complete.
Only together with the self-adjoint equation (\ref{D8}) one obtains a
full description for the Wigner operator. Therefore the covariant
version of the BGR equation is only one  part of the full covariant
transport theory. The question thus arises whether on the
three-dimensional level the BGR equation can be complete or not. To
answer this question we will reduce Eq.~(\ref{D8}) to the
three-dimensional level by performing an energy average, too, and
study its implications.

 \subsection{Energy average of full covariant theory}

The two self-adjoint equations (\ref{D7}) and (\ref{D8}) have a
complicated structure due to the occurrence of commutators and
anticommutators. But since they are equivalent to either equation
(\ref{D6}) or (\ref{D61}), we can consider equation (\ref{D6})
instead. To simplify the calculation further, we even remove the
$\gamma_0$-matrices and thus obtain the VGE \cite{Re11} equation:
 \begin{equation}
 \label{V1}
   \left[ \gamma^\mu \left(\Pi_\mu+\half i\hbar D_\mu \right)
          -m \right] \hat W_4(x,p) = 0 \, .
 \end{equation}
As in the scalar case, we will from now on consider the
electromagnetic field as classical and pass from the Wigner operator
to the Wigner function by leaving off the hats.

The Wigner function in spinor electrodynamics is a complex $4\times 4$
matrix. VGE discussed its spinor decomposition and derived a set of
coupled equations for the components of $W_4$. Because of their
characteristic transformation properties under Lorentz
transformations, it is convenient to choose the 16 matrices
$\Gamma_i=\{1,\ i\gamma_5, \ \gamma_\mu, \ \gamma_\mu\gamma_5, \
{1\over 2}\sigma_{\mu\nu}\}$ as the basis for an expansion of the
Wigner function in spin space:
 \begin{equation}
 \label{V2}
    \!\!\!\!\!\!\!\!\!\!\!\!
    W_4(x,p) = \quarter \left[ F(x,p) + i \gamma_5 P(x,p)
             + \gamma_\mu V^\mu(x,p) + \gamma_\mu \gamma_5 A^\mu(x,p)
             + \half \sigma_{\mu\nu} S^{\mu\nu}(x,p)\right] \, .
 \end{equation}
All the components $F, P, V_\mu, A_\mu,$ and $S_{\mu\nu}$ are
real functions since the basis elements $\Gamma_i$ transform under
hermitian conjugation like $W_4$ itself, $\Gamma_i^\dagger = \gamma_0
\Gamma_i \gamma_0$ (see Eq.~(\ref{D4})). They can thus
be interpreted as physical phase-space densities. The expansion
(\ref{V2}) decomposes the VGE equation into 5 coupled Lorentz
covariant equations for the spinor components. Since these components
are real and the operators $D$ and $\Pi$ are self-adjoint, one can
separate the real and imaginary parts of these 5 complex equations to
obtain 10 real equations:
 \begin{eqnarray}
 \label{V3}
    &&\Pi^\mu V_\mu = m F \, ,
 \nonumber \\
    &&\hbar D^\mu A_\mu = 2m P \, ,
 \nonumber \\
    &&\Pi_\mu F - \half \hbar D^\nu S_{\nu\mu}=m V_\mu \, ,
 \nonumber \\
    &&-\hbar D_\mu P+\epsilon_{\mu\nu\sigma\rho}\Pi^\nu S^{\sigma\rho}
      = 2m A_\mu \, ,
 \nonumber \\
    &&\half \hbar (D_\mu V_\nu-D_\nu V_\mu)
      + \epsilon_{\mu\nu\sigma\rho}\Pi^\sigma A^\rho
      = m S_{\mu\nu} \, ,
 \end{eqnarray}
and
 \begin{eqnarray}
 \label{V3a}
    &&\hbar D^\mu V_\mu = 0 \, ,
 \nonumber \\
    &&\Pi^\mu A_\mu =0 \, ,
 \nonumber \\
    &&\half \hbar D_\mu F = - \Pi^\nu S_{\nu\mu} \, ,
 \nonumber \\
    &&\Pi_\mu  P= -\quarter \hbar \epsilon_{\mu\nu\sigma\rho} D^\nu
      S^{\sigma\rho} \, ,
 \nonumber \\
    &&\Pi_\mu V_\nu-\Pi_\nu V_\mu = \half \hbar
      \epsilon_{\mu\nu\sigma\rho} D^\sigma A^\rho  \, .
 \end{eqnarray}

Peforming an energy average of the expansion of $W_4$, (\ref{V2}), we
obtain a full mapping from the four- to the three-dimensional
components introduced in \cite{Re12},
 \begin{eqnarray}
 \label{V4}
    W_3(x,{\bf p}) =
    && \!\!\!\!\!\!
       \quarter \Bigl[ f_0(x,{\bf p}) + \gamma_5 f_1(x,{\bf p})
                        -i \gamma_0 \gamma_5 f_2(x,{\bf p})
                        + \gamma_0 f_3(x,{\bf p})
 \\
   && + \gamma_5 \boldgamma{\cdot}{\bf g}_0(x,{\bf p})
      + \gamma_0 \boldgamma{\cdot}{\bf g}_1(x,{\bf p})
      - i\boldgamma{\cdot}{\bf g}_2(x,{\bf p})
      - \gamma_5 \boldgamma{\cdot}{\bf g}_3(x,{\bf p})\Bigr]\, ,
 \nonumber
 \end{eqnarray}
namely,
 \begin{eqnarray}
 \label{V5}
    && f_0(x,{\bf p})=\int dp_0 \, V_0(x,p) \, ,
 \nonumber \\
    && f_1(x,{\bf p})=-\int dp_0 \, A_0(x,p) \, ,
 \nonumber \\
    && f_2(x,{\bf p})=\int dp_0 \, P(x,p) \, ,
 \nonumber \\
    && f_3(x,{\bf p})=\int dp_0 \, F(x,p) \, ,
 \nonumber \\
    && {\bf g}_0 (x,{\bf p})=-\int dp_0 \, {\bf A}(x,p) \, ,
 \nonumber \\
    && {\bf g}_1 (x,{\bf p})=\int dp_0 \, {\bf V}(x,p) \, ,
 \nonumber \\
    && g_2^i (x,{\bf p})=-\int dp_0 \, S^{0i}(x,p) \, , \quad
       i=1,2,3\, ,
 \nonumber \\
    && g_3^i (x,{\bf p}) = \half \epsilon^{ijk}\int dp_0 \,
       S_{jk}(x,p) \, , \quad i=1,2,3 .
 \end{eqnarray}
As shown in \cite{Re12,Re18}, the physically interesting densities
such as charge current, energy momentum tensor and angular momentum
tensor, can also be expressed in terms of these spinor components. For
example, $f_0$ and ${\bf g}_0$ are the charge density and the spin density,
respectively.

The VGE equations (\ref{V3},\ref{V3a}) can be divided into two groups.
Group I contains the operator $\Pi_0(x,p)$ which involves $p_0$; group
II contains no $p_0$-dependence except for the one in the Wigner
function itself. For group I the energy average is straightforward.
The result is in complete agreement with the spinor decomposition of
the BGR equation \cite{Re12}:
 \begin{eqnarray}
 \label{BGR}
   &&\hbar(D_t f_0+{\bf D}{\cdot}{\bf g}_1)=0 \, ,
 \nonumber \\
   &&\hbar(D_t f_1+{\bf D}{\cdot}{\bf g}_0)=-2m f_2  \, ,
 \nonumber \\
   &&\hbar D_t f_2+2{\bf \Pi}{\cdot}{\bf g}_3=2m f_1  \, ,
 \nonumber \\
   &&\hbar D_t  f_3-2{\bf \Pi}{\cdot}{\bf g}_2)=0 \, ,
 \nonumber \\
   &&\hbar(D_t {\bf g}_0+{\bf D} f_1)-2{\bf \Pi}\times {\bf g}_1=0 \, ,
 \nonumber \\
   &&\hbar(D_t {\bf g}_1+{\bf D} f_0)-2{\bf \Pi}\times {\bf g}_0
     = -2m {\bf g}_2 \, ,
 \nonumber \\
   &&\hbar(D_t {\bf g}_2+{\bf D}\times {\bf g}_3)+2{\bf \Pi} f_3
     =2m {\bf g}_1 \, ,
 \nonumber \\
   &&\hbar(D_t {\bf g}_3-{\bf D}\times {\bf g}_2)-2{\bf \Pi} f_2 =0 \, .
 \end{eqnarray}

Due to the additional $p_0$-dependence from the operator $\Pi_0$, the
equations in group II can not be completely reduced to expressions
from the set of three-dimensional components given in Eq.~(\ref{V5}).
They contain additionally higher $p_0$-moments of the four-dimensional
components,
 \begin{eqnarray}
 \label{V6}
    &&\int dp_0\, p_0\, V_0 -{\bf \Pi}\cdot {\bf g}_1
      + \tilde\Pi_0 f_0=m f_3 \, ,
 \nonumber \\
    &&\int dp_0\, p_0\, A_0 +{\bf \Pi}\cdot {\bf g}_0
      - \tilde\Pi_0 f_1 = 0 \, ,
 \nonumber \\
    &&\int dp_0\, p_0\, P + \half \hbar {\bf D}\cdot {\bf g}_3
      + \tilde\Pi_0 f_2 = 0 \, ,
 \nonumber \\
    &&\int dp_0\, p_0\, F - \half \hbar {\bf D}\cdot {\bf g}_2
      + \tilde\Pi_0 f_3=m f_0 \, ,
 \nonumber \\
    &&\int dp_0\, p_0\, {\bf A} + \half \hbar {\bf D}\times {\bf g}_1
      + {\bf \Pi} f_1 - \tilde\Pi_0 {\bf g}_0 = -m {\bf g}_3 \, ,
 \nonumber \\
    &&\int dp_0\, p_0\, {\bf V} - \half \hbar {\bf D}\times {\bf g}_0
      - {\bf \Pi} f_0 + \tilde\Pi_0 {\bf g}_1 = 0 \, ,
 \nonumber\\
    &&\int dp_0\, p_0\, S^{0i}{\bf e}_i - \half \hbar {\bf D} f_3
      + {\bf \Pi}\times {\bf g}_3 - \tilde\Pi_0 {\bf g}_2 = 0 \, ,
 \nonumber \\
    &&\int dp_0\, p_0\, S_{jk} \epsilon^{jki}{\bf e}_i
      - \hbar{\bf D} f_2 +2{\bf \Pi}\times {\bf g}_2
      + 2 \tilde\Pi_0 {\bf g}_3 =2m {\bf g}_0 \, ,
 \end{eqnarray}
where we introduced the three-dimensional operator
 \begin{equation}
 \label{V7}
    \tilde \Pi_0 (x,{\bf p}) = i e \hbar
    \int^{1\over 2}_{-{1\over 2}}ds\, s\, {\bf E}({\bf x}
         + i s \hbar {\bf \nabla}_p,t){\cdot}{\bf \nabla}_p \, .
 \end{equation}
This second set of coupled equations for the components of $W_3$ does
not include the operator $D_t$. It corresponds to the energy average
of the spinor decomposition of the second self-adjoint equation
(\ref{D8}), where $D_t$ drops out due to the commutator in the first
term. These equations are therefore constraint equations rather than
equations of motion. They arise in addition to the BGR equations.

 \subsection{Classical limit}

Before discussing the non-perturbative consequences of Eqs.~(\ref{V6})
we show how they constrain the transport equations in the classical
limit. As $\hbar\to 0$, the original VGE equation can be
written in the following quadratic form:
 \begin{equation}
 \label{Cl1}
   (p^2-m^2) \, W_4(x,p) = 0 \, .
 \end{equation}
This shows that the classical Wigner operator and hence all its spinor
components are all on the mass shell, and thus they have positive and
negative energy solutions similar to Eq.(\ref{Sw5}),
 \begin{eqnarray}
 \label{Cl10}
    && f_i^{+(0)}(x,{\bf p})\,\delta(p_0-E_p),\quad
       f_i^{-(0)}(x,{\bf p})\,\delta(p_0+E_p)\, ,
 \nonumber\\
    && {\bf g}_i^{+(0)}(x,{\bf p})\,\delta(p_0-E_p),\quad
       {\bf g}_i^{-(0)}(x,{\bf p})\,\delta(p_0+E_p)\, ,
 \nonumber\\
    &&i=0,1,2,3\, .
 \end{eqnarray}
With this the remaining $p_0$-integrals in Eqs.~(\ref{V6}) can be done,
and everything can be fully expressed in terms of the
three-dimensional functions (\ref{V5}). Eqs.~(\ref{V6}) constribute
altogether six independent constraints for the positive and negative
energy parts of the spinor components:
 \begin{eqnarray}
 \label{Cl2}
   && f^{\pm (0)}_1=\pm {{\bf p}{\cdot}{\bf g}^{\pm (0)}_0 \over E_p} \, ,
 \nonumber\\
   && f^{\pm (0)}_2=0 \, ,
 \nonumber\\
   && f^{\pm (0)}_3=\pm {m\over E_p}\, f^{\pm (0)}_0 \, ,
 \nonumber\\
   && {\bf g}^{\pm (0)}_1=\pm {{\bf p}\over E_p}\, f^{\pm (0)}_0 \, ,
 \nonumber\\
   && {\bf g}^{\pm (0)}_2={{\bf p}\times {\bf g}^{\pm (0)}_0\over m} \, ,
 \nonumber\\
   && {\bf g}^{\pm (0)}_3=\pm {E_p^2 {\bf g}^{\pm (0)}_0
      -({\bf p}{\cdot}{\bf g}^{\pm (0)}_0){\bf p}\over m E_p} \, .
 \end{eqnarray}
In addition, the classical limit \cite{Re18} of the BGR equations in
(\ref{BGR}) gives another four constraints. Three of them are,
however, already included in (\ref{Cl2}), namely, the equations for
$f^{+(0)}_2, {\bf g}^{+(0)}_1$ and ${\bf g}^{+ (0)}_2$.  The last one,
$f^{+(0)}_1 = {\bf p}{\cdot}{\bf g}^{+(0)}_3/m$, is not independent of
the above equations either, but can be obtained through a linear
combination of Eqs.~(\ref{Cl2}). Therefore, Eqs.~(\ref{Cl2}) provide a
complete set of constraint equations for the three-dimensional spinor
components in the classical limit. There remain only two independent
components, for example, the charge density $f^{(0)}_0$ and the spin density
${\bf g}^{(0)}_0$. The classical limit of the BGR equations
yields only the positive energy parts of the latter 4 constraints,
and misses the first and the third additional equations from (\ref{Cl2}).

The classical transport equations for $f^{(0)}_0$ and ${\bf g}^{(0)}_0$
arise from the first order in $\hbar$ of the kinetic equations. From
the BGR equations in (\ref{BGR}), we have
 \begin{eqnarray}
 \label{Cl3}
   && (\partial_t + e {\bf E}{\cdot}{\bf \nabla}_p) f^{\pm(0)}_0
       + ({\bf \nabla} + e {\bf B}\times {\bf \nabla}_p)
         \cdot {\bf g}^{\pm(0)}_1 = 0 \, ,
 \nonumber\\
   && (\partial_t + e {\bf E}{\cdot}{\bf \nabla}_p) {\bf g}^{\pm(0)}_3
       - ({\bf \nabla} + e {\bf B}\times {\bf \nabla}_p)
         \times {\bf g}^{\pm(0)}_2
 \nonumber\\
   && \qquad\qquad + {{\bf p}\over m}
      \Big((\partial_t +e{\bf E}{\cdot}{\bf \nabla}_p) f^{\pm(0)}_1
          +({\bf \nabla}+e{\bf B}\times
            {\bf \nabla}_p){\cdot}{\bf g}^{\pm(0)}_0 \Big) = 0 \, .
 \end{eqnarray}
Using the constraint equations, some straightforward manipulations
lead to the following decoupled Vlasov-type
equation for the charge density,
 \begin{equation}
 \label{Cl4}
    \partial_t f^{\pm(0)}_0 \pm {\bf v}{\cdot}{\bf \nabla} f^{\pm(0)}_0
    + e ({\bf E} \pm {\bf v}\times {\bf B}){\cdot}{\bf \nabla}_p
         f^{\pm(0)}_0
    = 0 \, ,
 \end{equation}
 and the three-dimensional kinetic equation for the spin density,
 \begin{eqnarray}
 \label{Cl5}
   &&\partial_t {\bf g}^{\pm(0)}_0
     \pm ({\bf v}{\cdot}{\bf \nabla}) {\bf g}^{\pm(0)}_0
     + e \Big[({\bf E}\pm{\bf v}\times {\bf B}){\cdot}{\bf \nabla}_p \Big]
              {\bf g}^{\pm(0)}_0
 \nonumber\\
   &&\qquad\qquad  - {e\over E_p^2} \Big[
               ({\bf p}{\cdot}{\bf g}^{\pm(0)}_0){\bf E}
                -({\bf E}{\cdot}{\bf p}) {\bf g}^{\pm(0)}_0
                \Big]
            \pm {e\over E_p}{\bf B}\times {\bf g}^{\pm(0)}_0 = 0 \, .
 \end{eqnarray}

Through the energy averaging method, it is easy to prove that the above
equation is just the three-dimensional formulation of the covariant
transport equation for the classical axial vector $A_\mu^{(0)}$,
 \begin{equation}
 \label{BMT1}
    p^\mu(\partial_\mu-eF_{\mu\nu}\partial_p^\nu)A_\sigma^{(0)}
    =eF_{\sigma\rho}A^{\rho (0)} \, ,
 \end{equation}
which can be derived from the first order in $\hbar$ of the linear equations
(\ref{V3}) and (\ref{V3a}). From the discussions in \cite{Re11},
this transport
equation for $A_\mu^{(0)}$ is equivalent to the covariant
Bargmann-Michel-Telegdi (BMT) equation \cite{Re5,Re11,BMT} for a spinning
particle in a constant external field,
 \begin{equation}
 \label{BMT2}
    m{ds^\mu\over d\tau} = eF^{\mu\nu}(\tau)s_\nu(\tau) \, ,
 \end{equation}
where $s_\mu = A_\mu^{(0)}/(A^{(0)}{\cdot} A^{(0)})^{1/2}$ is the covariant
spin
phase-space density. Therefore, the kinetic equation (\ref{Cl5}) for the
spin density ${\bf g}^{\pm(0)}_0$ can be
recognized as the three-dimensional BMT equation. It describes the
procession of the spin-polarization in a homogeneous  external
electromagnetic field.

Equations (\ref{Cl4}) and (\ref{Cl5})
can be further simplified by introducing classical
particle and antiparticle distribution functions in an analogous way
to the case of scalar QED:
 \begin{eqnarray}
 \label {Cl6}
  f_i(x,{\bf p})
    &=& f_i^{+(0)}(x,{\bf p}) \, ,
  \nonumber\\
  {\bf g}_i(x,{\bf p})
    &=& {\bf g}_i^{+(0)}(x,{\bf p}) \, ,
  \nonumber\\
  \bar f_i(x,{\bf p})
    &=& f_i^{-(0)}(x,-{\bf p}) \, ,
  \nonumber\\
  \bar {\bf g}_i(x,{\bf p})
    &=& {\bf g}_i^{-(0)}(x,-{\bf p}) \, .
 \end{eqnarray}
The resulting equations for the particle distributions read
 \begin{equation}
 \label{Cl7}
    \partial_t f_0 + {\bf v}{\cdot}{\bf \nabla} f_0
    + e ({\bf E}+{\bf v}\times {\bf B}){\cdot}{\bf \nabla}_p f_0
    = 0 \, ,
 \end{equation}
 \begin{equation}
 \label {Cl8}
    \partial_t{\bf g}_0 + ({\bf v}{\cdot}{\bf \nabla}){\bf g}_0
    + e \Big[ ({\bf E}+{\bf v}\times {\bf B}){\cdot}{\bf \nabla}_p
        \Big] {\bf g}_0
    - {e\over E_p^2} \Big[ ({\bf p}{\cdot}{\bf g}_0){\bf E}
                          -({\bf E}{\cdot}{\bf p}){\bf g}_0
                     \Big]
    + {e\over E_p} {\bf B}\times {\bf g}_0 = 0 \, ;
 \end{equation}
those for the antiparticle distributions $\bar f_0$ and $\bar {\bf
g}_0$ differ only by a minus sign in front of the electric charge $e$.

The other particle and antiparticle distribution functions can be
derived from the constraints (\ref{Cl2}):
 \begin{eqnarray}
 \label {Cl12}
   &&     f_1 = {{\bf p}{\cdot}{\bf g}_0\over E_p}\, , \qquad
     \bar f_1 = {{\bf p}{\cdot}{\bar{\bf g}}_0 \over E_p} \, ,
 \nonumber\\
   &&     f_2 = \bar f_2 = 0 \, ,
 \nonumber\\
   &&     f_3 = {m\over E_p}f_0 \, , \qquad
     \bar f_3 = -{m\over E_p} \bar f_0 \, ,
 \nonumber\\
   &&   {\bf g}_1 = {{\bf p}\over E_p}f_0\, , \qquad
   \bar {\bf g}_1 = {{\bf p}\over E_p}\bar f_0 \, ,
 \nonumber\\
   &&   {\bf g}_2 = {{\bf p}\times {\bf g}_0\over m}\, , \qquad
   \bar {\bf g}_2 = - {{\bf p}\times \bar {\bf g}_0\over m} \, ,
 \nonumber\\
   &&   {\bf g}_3 = {E_p^2{\bf g}_0 - ({\bf p}{\cdot}{\bf g}_0){\bf p}
                     \over E_p m} \, , \qquad
   \bar {\bf g}_3 = -{E_p^2 \bar {\bf g}_0
                      - ({\bf p}{\cdot}{\bar{\bf g}}_0){\bf p}
                     \over E_p m} \, .
 \end{eqnarray}

The discussion above demonstrates the incompleteness of the equal-time
BGR formulation, (\ref{D9}) and (\ref{BGR}), in the classical limit.
Without the additional constraints (\ref{V6}) one has \cite{Re18} $4$
independent components, $f^{(0)}_0, f^{(0)}_3, {\bf g}^{(0)}_0$ and
${\bf g}^{(0)}_3$, and one obtains two groups of coupled equations,
where the first one couples $f^{(0)}_0$ to $f^{(0)}_3$ and the second
one couples ${\bf g}^{(0)}_0$ to ${\bf g}^{(0)}_3$. The
additional constraints arising from the complete covariant formulation
allow to decouple $f^{(0)}_0$ from $f^{(0)}_3$ and ${\bf g}^{(0)}_0$
from ${\bf g}^{(0)}_3$. Therefore, we need only one scalar and one
vector component to completely describe the classical behavior of the
Wigner operator. In fact, already VGE \cite{Re11} pointed out in their
covariant formulation that to any finite order in $\hbar$ the
pseudoscalar $P$, vector $V^\mu$ and antisymmetric tensor
$S^{\mu\nu}$
components can be expressed in terms of the scalar $F$ and axial
vector $A^\mu$ components. Furthermore, the general relationship
$\Pi_\mu A^\mu = 0$ in (\ref{V3}) shows that $A_0$ is not an
independent component either. Thus only 4 of the $16$ components of
the Wigner function are dynamically independent. Through our mapping
between the four- and three-dimensional components, which results
from taking the energy average, we are thus able to construct
transport equations with only $2$ independent densities,
the scalar
mass density $f_3$ and the vector spin density ${\bf g}_0$, to any
order of the semiclassical expansion.

 \subsection{General kinetic equations in three-dimensional form}

In this subsection, we go beyond the semiclassical expansion and study
the full quantum kinetic equations in three-dimensional form. This is
important for the treatment of non-perturbative processes like pair
creation in strong electric fields. Beyond the classical limit, the
classical mass-shell condition is generally violated. The quantum
Wigner function is no longer a $\delta$-function in $p_0$ located at
the mass-shell energy, and for the elimination of the higher
$p_0$-moments of the covariant spinor components from Eqs.~(\ref{V6})
we must follow a different path. Following the procedure from Section
2.3.1, we multiply the equations in group I of (\ref{V3},\ref{V3a}) by
$p_0$ from the left and then take the energy average of these
equations. We find
 \begin{eqnarray}
 \label{G1}
   &&D_t\int dp_0\, p_0\, V_0 + {\bf D}{\cdot}\int dp_0\, p_0\, {\bf V}
     + I f_0 + {\bf J}{\cdot}{\bf g}_1 = 0 \, ,
 \nonumber \\
   &&D_t\int dp_0\, p_0\, A_0 + {\bf D}{\cdot}\int dp_0\, p_0\, {\bf A}
     - I f_1 - {\bf J}{\cdot}{\bf g}_0 = 2 m \int dp_0\, p_0\, P \, ,
 \nonumber \\
   &&D_t\int dp_0\, p_0\, P + I f_2 + 2\, {\bf K}{\cdot}{\bf g}_3
     = - {\bf \Pi}{\cdot}\int dp_0\, p_0\, \epsilon^{ijk} S_{jk} {\bf e}_i
       - 2 m \int dp_0\, p_0\, A_0 \, ,
 \nonumber \\
   &&\half \Big[ D_t \int dp_0\, p_0\, F + I f_3
      -2 \, {\bf K}{\cdot}{\bf g}_2 \Big] =
      - {\bf \Pi}{\cdot}\int dp_0\, p_0\, S^{0i}{\bf e}_i \, ,
 \nonumber \\
   &&D_t \int dp_0\, p_0\, {\bf A} + {\bf D} \int dp_0\, p_0\, A_0
     - I {\bf g}_0 - {\bf J} f_1 +2 \, {\bf K}{\times}{\bf g}_1
     = 2\, {\bf \Pi}\times \int dp_0\, p_0\, {\bf V} \, ,
 \nonumber \\
   &&\half \Big[ D_t \int dp_0\, p_0\, {\bf V}
     + {\bf D} \int dp_0\, p_0\, V_0 + I {\bf g}_1 + {\bf J} f_0
     - 2\, {\bf K}{\times}{\bf g}_0 \Big] =
 \nonumber\\
   &&\qquad\qquad\qquad\qquad\qquad\qquad\qquad\qquad
     - {\bf \Pi}{\times}\int dp_0\, p_0\, {\bf A}
     + m \int dp_0\, p_0\, S^{0i}{\bf e}_i \, ,
 \nonumber \\
   &&D_t \int dp_0\, p_0\, S^{0i}{\bf e}_i
     - {1\over 2} {\bf D}{\times}\int dp_0\, p_0\,
        \epsilon^{ijk} S_{jk}{\bf e}_i
     - I {\bf g}_2 - {\bf J}{\times}{\bf g}_3 - 2\, {\bf K} f_3 =
 \nonumber\\
   &&\qquad\qquad\qquad\qquad\qquad\qquad\qquad\qquad
     2\, {\bf \Pi}\int dp_0\, p_0\, F
     - 2 m \int dp_0\, p_0\, {\bf V} \, ,
 \nonumber\\
   &&\half \left[ \half D_t \int dp_0\, p_0\,
                        \epsilon ^{ijk} S_{jk} {\bf e}_i
      + {\bf D}{\times}\int dp_0\, p_0\, S^{0i}{\bf e}_i
      + I {\bf g}_3 - {\bf J}{\times}{\bf g}_2 - 2\, {\bf K} f_2\right] =
 \nonumber\\
   &&\qquad\qquad\qquad\qquad\qquad\qquad\qquad\qquad
     {\bf \Pi}\int dp_0\, p_0\, P \, ,
 \end{eqnarray}
with
 \begin{eqnarray}
 \label{G2}
   &&I(x,{\bf p}) = i e \int^{1\over 2}_{-{1\over 2}} ds\, s\,
     (\partial_t{\bf E})({\bf x}+is{\bf \nabla}_p,t){\cdot}{\bf \nabla}_p
   \, ,
 \nonumber \\
   &&{\bf J}(x,{\bf p}) = i e \int^{1\over 2}_{-{1\over 2}} ds\, s\,
     (\partial_t{\bf B})({\bf x}+is{\bf \nabla}_p,t)
      \times{\bf \nabla}_p - e \int^{1\over 2}_{-{1\over 2}} ds\,
      {\bf E}({\bf x}+is{\bf \nabla}_p,t) \, ,
 \nonumber \\
   &&{\bf K}(x,{\bf p}) = e \int^{1\over 2}_{-{1\over 2}} ds\, s^2\,
      (\partial_t{\bf B})({\bf x}+is{\bf \nabla}_p,t)
       \times {\bf \nabla}_p + i e \int^{1\over 2}_{-{1\over 2}} ds\, s\,
       {\bf E}({\bf x}+is{\bf \nabla}_p,t) \, .
 \end{eqnarray}

We now combine Eqs.~(\ref{V6}) and (\ref{G1}) and eliminate all higher
$p_0$-moments. In order to arrive at a set of independent contraints
we also use the BGR equations and remove all information already
contained in them. After a straightforward but tedious calculation we
finally derive the following constraints:
 \begin{eqnarray}
 \label {G3}
    &&L f_0+{\bf M}\cdot {\bf g}_1=0 \, ,
 \nonumber \\
    &&L f_1+{\bf M}\cdot {\bf g}_0=0 \, ,
 \nonumber \\
    &&L f_2+2{\bf N}\cdot {\bf g}_3=0 \, ,
 \nonumber \\
    &&L f_3-2{\bf N}\cdot {\bf g}_2=0 \, ,
 \nonumber \\
    &&L {\bf g}_0 -{\bf M} f_1 -2{\bf N}\times {\bf g}_1 = 0 \, ,
 \nonumber \\
    &&L {\bf g}_1 +{\bf M} f_0 +2{\bf N}\times {\bf g}_0 =0 \, ,
 \nonumber \\
    &&L {\bf g}_2 +{\bf M}\times {\bf g}_3 -2{\bf N} f_3 = 0 \, ,
 \nonumber \\
    &&L {\bf g}_3 -{\bf M}\times {\bf g}_2 +2{\bf N} f_2 = 0 \, ,
 \end{eqnarray}
with
 \begin{eqnarray}
 \label{G4}
    && \!\!\!\!\!\!\!\!\!\!\!
      L(x,{\bf p}) = i e \int^{1\over 2}_{-{1\over 2}} ds\, s\,
      \Big({\bf \nabla}\times {\bf B}({\bf x}+is{\bf \nabla}_p,t)
      \Big) \cdot {\bf \nabla}_p \, ,
 \nonumber \\
    && \!\!\!\!\!\!\!\!\!\!\!
      {\bf M}(x,{\bf p}) = i e \int^{1\over 2}_{-{1\over 2}} ds\, s\,
      {\bf \nabla} \Big({\bf E}({\bf x}+is{\bf \nabla}_p,t){\cdot}{\bf
          \nabla}_p\Big) + e \int^{1\over 2}_{-{1\over 2}} ds\,
      \Big({\bf E}({\bf x}+is{\bf \nabla}_p,t) - {\bf E}(x)\Big) \, ,
 \nonumber \\
    && \!\!\!\!\!\!\!\!\!\!\!
       {\bf N}(x,{\bf p}) = \quarter e \int^{1\over 2}_{-{1\over 2}}ds\,
       ({\bf \nabla}_p{\cdot}{\bf \nabla})\,
       {\bf E}({\bf x}+is{\bf \nabla}_p,t) + {\bf K}(x,{\bf p}) \, .
 \end{eqnarray}

{\em The complete set of kinetic equations for the Wigner function
$W_3(x,{\bf p})$ (resp. its spinor components) is given by the BGR
equations (\ref{BGR}) and the constraints (\ref{G3}).} These equations
together form the complete transport theory in its three-dimensional
formulation. They are fully equivalent to the covariant kinetic
approach defined by either of the two equations (\ref{D6}) or by the
pair of self-adjoint equations (\ref{D7},\ref{D8}). While the
equations in (\ref{G3}) do not involve time derivatives and thus do
not by themselves describe transport, they provide essential
constraints for the Wigner operator.

The constraints (\ref{G3}) hold for arbitrary external fields. For
homogeneous fields, the spatial derivatives of the electric and
magnetic fields ${\bf E}$ and ${\bf B}$ in the operators $L, {\bf M}$
and ${\bf N}$ vanish, and the constraints (\ref{G3}) disappear
identically. This shows that the equal-time BGR theory is in fact
complete for the case of a spatially constant external field, i.e. for
the case studied in \cite{Re12}. The application of the BGR theory in
\cite{Re12} to the problem of pair creation in spatially homogeneous
but time-dependent electric fields is therefore safe.

 \section{Conclusions}

Relativistic kinetic theory, formulated in terms of gauge covariant
Wigner operators, provides a useful method to study transport problems
in quantum theory, especially the phase-space evolution of high
temperature and density plasmas. There exist two paths to formulate
such a transport theory, starting from 4-dimensional or from
3-dimensional momentum space, respectively. In this paper we have
shown a way to connect these two approaches directly and arrived
at the 3-dimensional formulation by taking an energy average of the
covariant 4-dimensional formulation.

We demonstrated the method explicitly for scalar and spinor QED with
external electromagnetic fields. In scalar QED, our approach started
directly from the Klein-Gordon equation and thus avoided the
complications connected with the $2\times 2$ matrix structure of the
Feshbach-Villars representation \cite{Re13}. Using the energy
averaging method and taking the semiclassical limit, we directly
arrived at the classical Vlasov equation for on-shell particles. We
pointed out that there are no first-order quantum corrections due to
the absence of spin. This may partially explain the success of the
classical Vlasov approach as a good approximation for the description
of many quantum systems involving scalar fields \cite{Re22}.

Beyond the semiclassical expansion, we could still perform the energy
average analytically for the case of spatially constant
electromagnetic fields, and we recovered the correct result for the
pair creation rate in a homogeneous electric field. Our calculation
was much easier than the one based on 3-dimensional transport theory
in the Feshbach-Villars representation, due to the Lorentz scalar
nature of the Wigner function employed by us.

In spinor electrodynamics, we concentrated on the question of
completeness of the 3-dimensional (BGR) transport theory. Employing
the energy averaging method, we identified a covariant version of the
equal-time BGR equation, but found that this covariant equation is not
equivalent to the original Dirac equation. Integrating instead the
covariant VGE equation, which forms a complete description for the
spinor Wigner function, over the energy $p_0$, we derived a set of
3-dimensional (equal-time) transport and constraint equations
which is complete and contains the BGR equations as a subset.
The additional equations can be understood as resulting from the
energy average of the 4-dimensional constraint or generalized
mass-shell condition. They cannot be obtained by the direct equal-time
approach of BGR. In the semiclassical limit they allow to reduce the
number of independent spinor components of the Wigner function from 8
in the BGR approach to 4, one scalar charge density and 3 vector
components of the spin density. The equations for the charge density and
for the spin density decouple, and each of them satisfies a
Vlasov-type transport equation in (6+1)-dimensional phase space. We
also derived the additional constraints explicitly for the full
quantum case and showed that they vanish only for homogeneous external
electromagnetic fields.

Either of the two formulations has its advantages and disadvantages.
The approach based on 4-dimensional momentum space (8-dimensional
phase space) is manifestly Lorentz covariant, but setting it up as an
initial value problem poses certain difficulties \cite{Re12}: since
the covariant Wigner function is a 4-dimensional Wigner transform of
the density matrix, its calculation at $t=-\infty$ requires knowledge
of the fields at all times. Since this knowledge does not a priori
exist, the covariant transport equations can only be solved with
phenomenologically motivated forms for the Wigner function in the far
past. The formulation in 3-dimensional momentum space (i.e.
(6+1)-dimensional phase space) does not have this problem: it requires
the density matrix only at equal times, and the Wigner function at
$t=-\infty$ can be directly calculated from the fields at $t=-\infty$.
Setting up the initial value problem is therefore straightforward in
this approach. For the calculation of the pair creation rate in
homogeneous electric fields this seems to be crucial: also our
calculation was based on the 3-dimensional version after taking the
energy average.

In both formulations the transport equations are supplemented by
constraint equations. In the covariant approach this is essentially
one $4\times 4$ matrix equation, and it is easily interpreted as a
generalized mass-shell condition. In the equal-time approach the
constraints are less transparent, and their derivation via the energy
averaging procedure is actually rather tedious. For homogeneous
external fields, however, they disappear to all orders in $\hbar$, so
for this particular case the equal-time approach has a decisive
advantage over the covariant approach where this does not happen. In
all other cases a large number of constraints, Eqs.~(\ref{G3}), have to
be solved together with the BGR kinetic equations (\ref{BGR}).

\vspace{0.25cm}
\noindent{\bf Acknowledgments} \\
P.Z. wishes to thank the Alexander von Humboldt Foundation
for a fellowship. U.H. would like to thank the participants of the ECT*
Workshop in Trento
(Oct. 1994) on Parton Production in the Quark-Gluon Plasma, in particular
J. Eisenberg, Y. Kluger and E. Mottola, for stimulating discussions
following a presentation of part of this work. We are also grateful to S. Ochs
for helpful remarks.
This work was supported by DFG, BMFT, and GSI.


\end{document}